\documentclass[12pt]{article} \textheight 23cm
\usepackage{graphicx}
\usepackage{amsmath}
\usepackage{color}
\begin{document} \begin{center}
{\huge \bf Fracture of  Composites: Simulation by a Spring Network Model  }\\ \vskip 0.5cm
Supti Sadhukhan$^1$, Tapati Dutta$^2$, Soma Nag$^3$
and Sujata Tarafdar$^3$\\
\vskip 0.5cm

$^1$ Physics Department, Jogesh Chandra Chaudhuri College,
\\ Kolkata 700033, India\\
$^2$ Physics Department, St. Xavier's College, \\Kolkata 700016, India\\
$^3$ Condensed Matter Physics Research Centre,\\Physics Department, Jadavpur University,\\ Kolkata 700032, India\\

\end{center}
\vskip 1cm
\noindent {\bf Abstract}\\
\noindent
Composite materials are often stronger than their constituents. We demonstrate this through a spring network model on a square lattice. Two different types of sites (A and B) are distributed randomly on the lattice, representing two different constituents. There are springs of three types connecting them (A-A, B-B and A-B). We assign two  spring parameters for each type of spring. these are a spring constant and a breaking threshold. Here we show that intermediate compositions may require higher energy to induce the first sample-spanning break than either pure A or pure B. So the breaking energy goes through a maximum as the concentration of one component varies from 0 to 100\%. The position and height of the peak depend on the spring parameters. Moreover, certain combinations of spring parameters can produce composites, which do not break upto a specified strain limit. Thus a fracture 'percolation threshold' may be defined.\\

\section{Introduction}

Properties of composite materials are often not simply linear combinations of the properties of the constituents. The strength of many composites exceeds the strength of either of the components \cite{gong,zuda}, even when one of the components is simply void space. In the present paper we suggest an explanation for this kind of strongly non-linear behavior using a simple spring network model in two dimensions. The model developed in \cite{sadhu} is related to the fracture models discussed in the review by \cite{nukala}, but is applied here to composites.

 A recent study on polymer-clay composites \cite{jpcm} used a spring network model to simulate crack patterns formed by desiccation. Experimental results agreed well with the simulations. A modification of the same model is employed here. 
The basic model is a two-dimensional network of springs on a square lattice. The nodes on the square lattice may be of two types - A and B, corresponding to two different species making up the composite. We designate the fraction of A sites by $Pr$. The springs connecting the sites can thus be of three types A-A, A-B and B-B. The springs are Hookean, having a definite spring constant up to a certain threshold strain, where the spring fails.  

So each type of spring is assigned two parameters - a spring constant (g) and a threshold breaking strain (s). Varying these six parameters it is possible to simulate different situations leading to new and interesting results. Several variants of this problem are possible. We look for the energy required to
break the sample into at least two disjoint pieces.

We assume all types of bonds -  A-A, A-B and B-B to have different but finite spring constants and breaking thresholds. We look at the first appearance of a sample spanning crack while varying $Pr$. We calculate the total energy required to produce this crack and show that it may go through a pronounced maximum as $Pr$ varies from 0 to 1. 

Another point of interest is the condition where a sample spanning crack fails to form up to the strain limit considered. The parameter values where this transition happens defines a 'fracture percolation threshold'.

The results presented here are averages over several configurations (typically 100), since the process is stochastic.

\section{The Spring Network Model}

In simulating the crack formation in the composite system, we utilize our code for a 2-dimensional spring network, which was developed in \cite{peel}. It is modified suitably for the present problem. The top view of the composite system is represented by a square lattice, with the nodes representing A or B particles. A fraction $Pr$ of the nodes are randomly assigned to be of type A, the remaining are of type B. Bonds between the nodes are assumed to be springs, which are Hookean up to a certain threshold strain, beyond this they break. Each type of spring is characterized by 2 parameters, a spring constant ($g(ij)$)and a breaking threshold ($s(ij)$), where i,j represent A or B. So there are 6 spring parameters in all.

Unbalanced forces start to act on the nodes as the system is strained. Here strain is implemented through a time dependent contraction of the springs. A molecular dynamics (MD)
approach has been used to calculate
the evolution of the system. As the natural length of the springs is reduced, the 4 neighboring springs cause a net force to act on each node.
We restrict the movement of the boundary layer springs, allowing motion only {\it along} the length of the boundary, but not normal to it.

An iterative scheme is utilized, which describes the successive decrease in natural length of
each spring with time. The same rule has been used earlier by Sadhukhan et al.\cite{sadhu} for desiccation cracking.
\begin{equation}d_{\tau+1}=d_{\tau}(1-b/r^\tau) \label{shrink} \end {equation} where $d_{\tau+1}$ and $d_\tau$ represent respectively the
natural length between two nodes in the $(\tau+1)$th and $\tau$th time step, $b$ is a constant and $r$ is a
parameter which controls the rate of decrease in the natural length from one time step to the
next. This empirical rule causes the shrinking rate to decrease gradually
with time. So the strain acts up to a certain point of time after which the shrinking of the springs becomes negligible. The final value of $d$ is $d_{min}$.

In \cite{peel}  eq.(\ref{shrink}) is integrated and given a form more amenable to
the present MD formulation. Eq.(\ref{shrink}) can be written as

\begin{equation} d_\tau = \exp (br^{ -\tau}/\ln r) \end{equation}
In the present paper we use the same form.

$d_\tau$ is normalized
to $d_0$ at $\tau$=0. 

In the first set of simulations, the parameters have been assigned values such that all springs break in the limit up to which the strain acts. These values are $b$ = 0.10 and $r$ = 1.02. 

The molecular dynamics proceeds as follows.
As the the bonds shrink, each node is acted on by a net force determined by shrinkage of the four adjacent nearest neighbor springs. Even for a pure  system, unbalanced forces arise due to the boundary condition imposed at the four sides. For $Pr \ne 1$ additional asymmetric forces arise due to the difference in parameters for the A-A, B-B and A-B type springs. So each node takes up a new position after time $\delta t$, determined by ${\bf v} \delta t$. This procedure is equivalent to a simplified form of Verlet's algorithm \cite{peel}. Here ${\bf v}$ is the instantaneous velocity of a particle calculated from  the net force (i.e. the acceleration).

We find that the optimum choice of $\delta t$ is  0.05 . After every time interval $\delta t$, the maximum force on a particle is noted. We then check whether the strain on any spring has exceeded the threshold, in which case it breaks.
If a number of springs cross the threshold simultaneously, the one with the highest strain
breaks. If again, there are more than one springs with the same highest strain,
one is randomly chosen to break. The next drying occurs in the next interval.

The molecular dynamics runs until $d_{min}$ is reached and there is no further shrinking.

We have simulated crack formation on a square network with sides 50 units long. However, a size $60 \times 60$ is employed where lower system sizes give erratic results. The fraction of A particles $Pr$ has been varied from 0 to 1 in intervals 0.1 or less. The parameters are designated as follows - $g(ij)$ represents the spring constant of the spring connecting sites of type i and j. Here i and j can be either A or B. The breaking threshold $s(ij)$ is the maximum strain the spring can withstand. If the threshold strain is exceeded, the spring breaks and the gap becomes part of a crack. 

A spanning crack is identified by a variant of the well-known algorithm given by Hoshen and Kopelman \cite{Hoshen}. A spanning fracture is formed if a crack starting
from any one side of the sample reaches the opposite side. A crack that begins on
one side and ends on an adjacent side, does not constitute a spanning cluster, by our present definition. We check for the first appearance of spanning fractures in either vertical or horizontal directions.

\section{Simulation Results}

It is good to keep in mind that A sites occur with probability $Pr$, so the fractional number of A-A, A-B and B-B bonds are respectively
\begin{equation}
 N(AA) = Pr^2
 ;N(AB) = 2Pr(1-Pr);
 N(BB) = (1-Pr)^2
\end{equation}

\begin{figure}[ht]
\begin{center}
\includegraphics[width=14.0cm, angle=0]{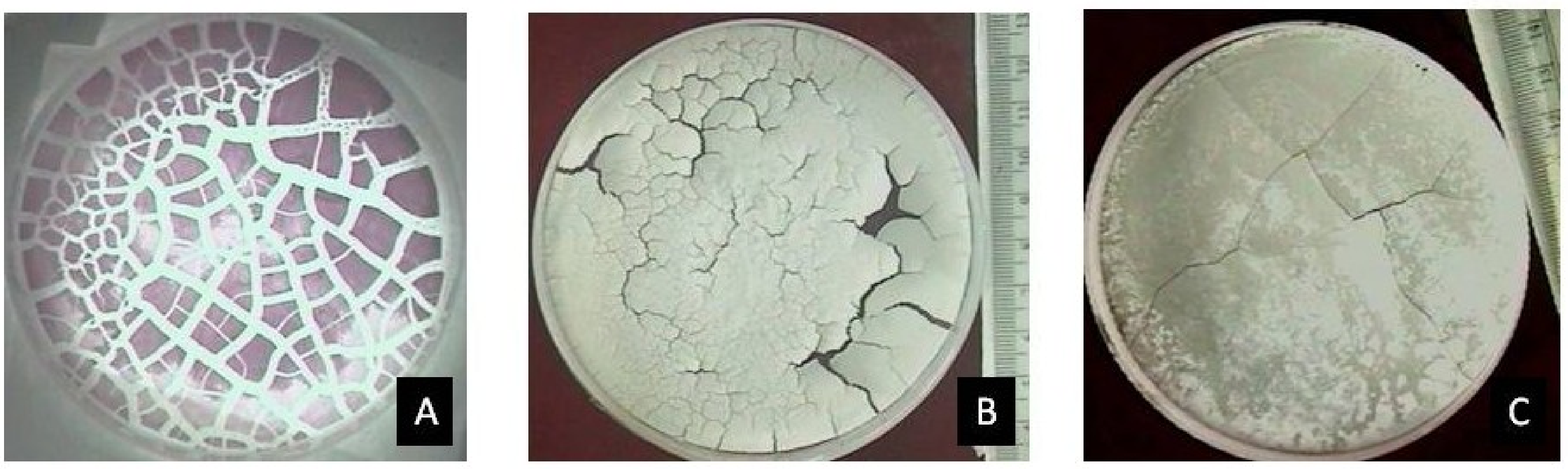}
\end{center}
\caption{Crack formation on drying suspensions of (A) laponite, (B) a 50-50 mixture of laponite and cement and (C) cement are shown. Methanol was used as the solvent. The cracks in (A) and (C) are straight while in (B) they are tortuous. } \label{cement}
\end{figure}

We find that in the simulated composites the cracks are highly tortuous, compared to the pure systems where they are straight. This is observed experimentally in real composites. 
We show in fig(\ref{cement}) crack patterns obtained on desiccating suspensions of cement and laponite in methanol. Preliminary results for pure cement, pure laponite and a 50-50 mixture are shown here. Experiments for different compositions are in progress.

In the simulation,we consider a composite with one component of low strength and another of moderate strength. We show that intermediate compositions may have higher breaking energy than each pristine component. Here all bonds are breakable but have different parameters.

If B-B bonds are weak and A-A bonds stronger, the pure systems when strained look similar, with nearly straight cracks. However, the energy spent in breaking the pure A system is larger. For a mixture of A and B, where cracks are tortuous and all types of bonds need to be broken to crack the system,  the total energy required for a spanning crack may be even more than that for the pure A sample.

We calculate the energy for fracture as follows: Assuming the springs to be Hookean until they break, the energy for breaking is the energy when breaking strain $s(ij)$ is reached. Since the spring constant $g(ij)$ is a measure of the elastic coefficient, we may write the energy for breaking a bond of type $ij$ as

\begin{equation}
 e(ij) = g(ij) s(ij)^2
\end{equation}
The total energy for the strained system is then
\begin{equation}
 E = \Sigma e(ij) N(ij)
\end{equation}

where $N(ij)$ is the number of broken bonds of type $ij$.

\begin{figure}[ht]
\begin{center}
\includegraphics[width=14.0cm, angle=270]{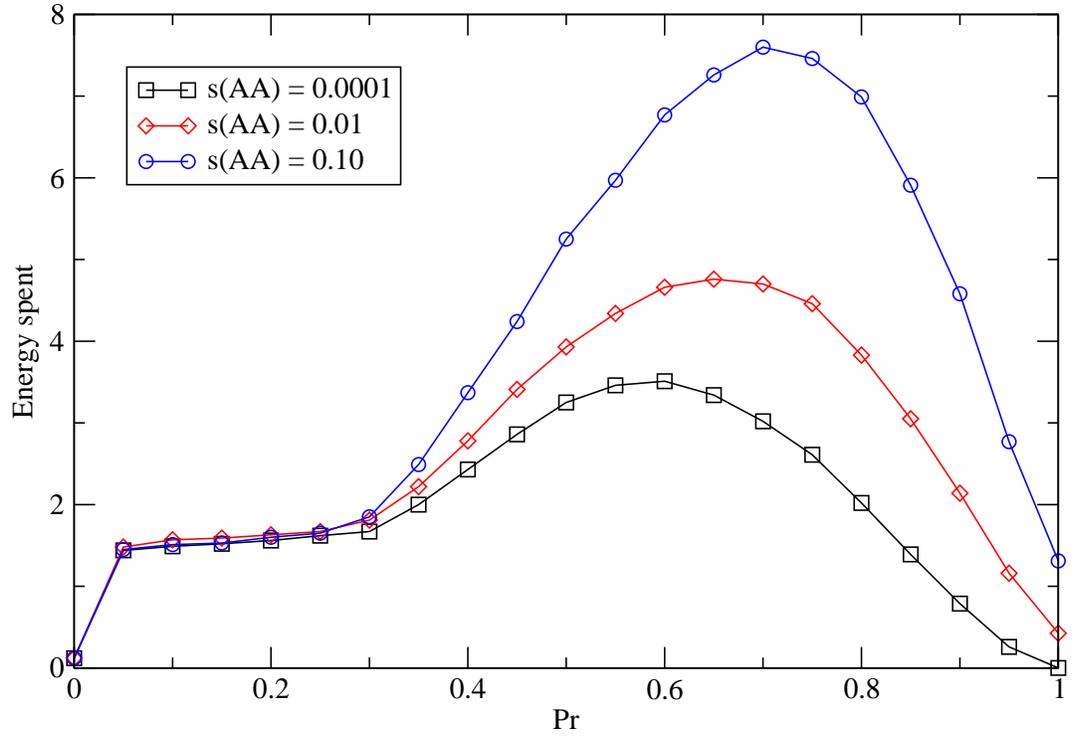}
\end{center}
\caption{Breaking energy for parameters $g(AA)=g(AB)$= 1.0, $g(BB)$=0.1, i.e. B is less stiff, $s(AB)=s(BB)$=0.1 and $s(AA)$ takes different values. There is a maximum in energy for intermediate $Pr$ and the height and position of the maximum shifts with $s(BB)$. } \label{venergy}
\end{figure}

We first show the results for the following set of parameters $g(AA)=g(AB)=1.0$, $g(BB)=0.1$, $s_{BB}=0.001$, $s_{AB}=s_{AA}=0.1$. Figure(\ref{venergy}) shows that
the total energy $E$ spent up to the point where the system just breaks into two, is more for $Pr$ between 0 and 1, compared to the values at either limit. 
 With this set of parameters, the maximum strength is at $Pr$ = 0.6, where the $E(0.6)$ is larger than the value at $Pr$ = 0, i.e. $E(0.0)$ by a factor of 20. So the pristine stronger component is much weaker than the composite formed with a material which is still weaker in the pristine state.
 It will be interesting to look for  other combinations of spring parameters where this character is observed.

  The visual appearance of the system at the breaking point for different values of $Pr$ is shown in figure(\ref{energymax}).
  \subsection{Effect of varying the spring parameters}
 One may ask how the position and the height of the peak in energy depend on the spring parameters. We calculate the energy at breaking for different values of $g(BB)$, with other parameters held constant. We fix $g(AA)=g(AB)=1.0$ and $g(BB)=0.1$.

 With $s(AA)=s(AB)=0.1$ we vary $s(BB)$. The energy required for the first break is shown in fig(\ref{venergy}). It is seen that the peak position and height both change with $s(BB)$. The number of broken bonds are however the same for all these
 cases, only the energy changes. The number of bonds of different type which must be broken to get a sample-spanning crack are shown in fig(\ref{bondbr}). 

Next we vary the parameter $s(AA)$, i.e. the breaking threshold of the stronger component. In the previous work \cite{supti}, the percolation aspect of the problem was explored, making one or
more spring unbreakable with $s(ij)$= $\infty$. Here we see that a transition to an unbroken
regime can be obtained with  finite thresholds as well. 
We start with $g(AA)=g(AB)=1.0$ and $g(BB)$=0.1, b=0.05, r= 1.10. So bonds between B-B are less stiff. 
Now we fix $s(AB)=s(AA)$=1.0 and allow $s(BB)$ to take on different values starting with a value equal to the other two $s$.
As $s(BB)$ increases from 0.1 to 1.0, we find a transition. For values up to 0.3, 100\% bonds break for pure B (i.e. $Pr$ = 0), a rapid transition follows and at $Pr$ = 0.05, we find no breaks, the spanning crack again appears after $Pr$ crosses 0.5. 

\begin{figure}[ht]
\begin{center}
\includegraphics[width=14.0cm, angle=0]{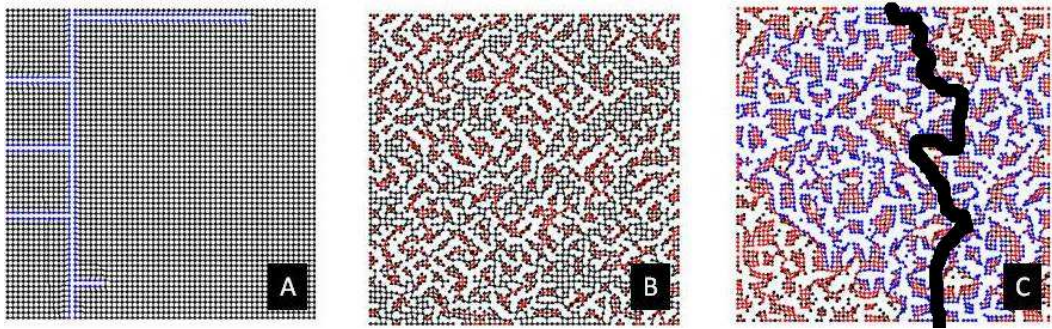}
\end{center}
\caption{ Snapshots for $s(AA)=s(AB)$=0.1, s(BB) = 0.5, $g(AA)=g(AB)$=1.0, $g(BB)$=0.1. $Pr$ values of 0.2, 0.6, 0.8, 0.9, 0.95 and 1.0 are shown in (A) - (F) respectively. A continuous spanning crack is present only in (C) shown in blue, the line shows one possible connecting path.} \label{pprfig}
\end{figure}

\begin{figure}[ht]
\begin{center}
\includegraphics[width=14.0cm, angle=0]{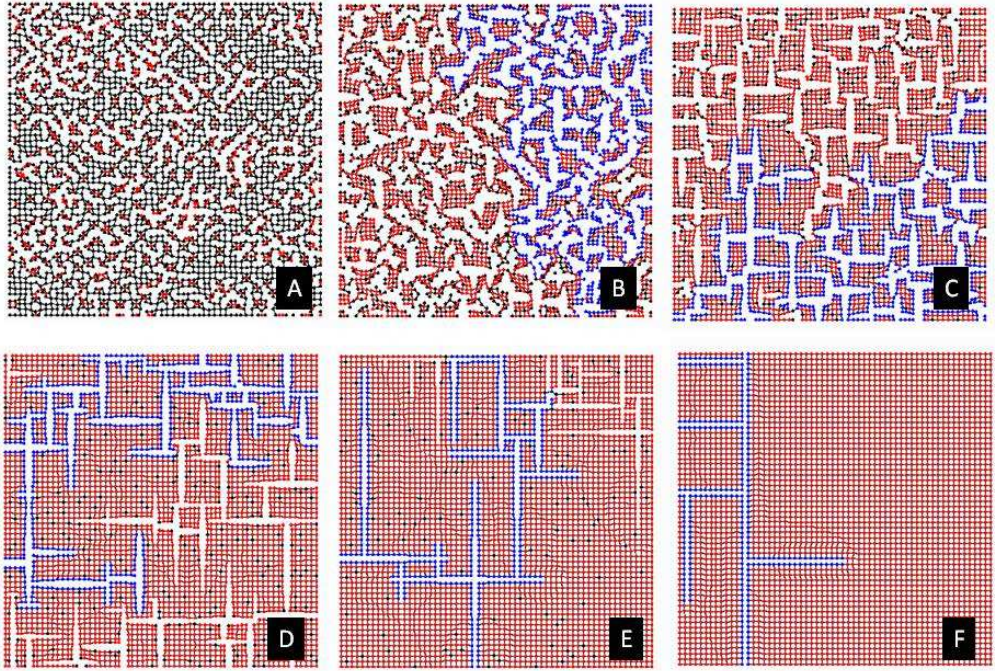}
\end{center}
\caption{Snapshots for $s(AA)=s(AB)= s(BB)$=0.1, , $g(AA)=g(AB)$=1.0, $g(BB)$=0.1. $Pr$ values of 0,0.2 and 0.6 are shown in (A), (B) and (C) respectively } \label{pprfig1}
\end{figure}

'Snapshots' of the system for $s(BB)$= 0.1 for $Pr$ = 0, 0.2 and 0.6 are shown respectively in fig(\ref{pprfig}) A to C.
For $s(BB)$ = 0.5 there are no spanning cracks up to $Pr$=0.55, but after that 100\% of the configurations break up, when s(BB) = 1.0 the appearance of the spanning crack shifts to 0.8. 
'Snapshots' are shown in fig(\ref{pprfig1}) with $Pr$ = 0.2, 0.6, 0.8, 0.9, 0.95 and 1.0 in A, B, C, D and E respectively.

These sets of data were simulated for system size = 60 $\times$ 60.  For lower sizes there are very strong fluctuations. This is natural near the critical regime.

\begin{figure}[ht]
\begin{center}
\includegraphics[width=14.0cm, angle=270]{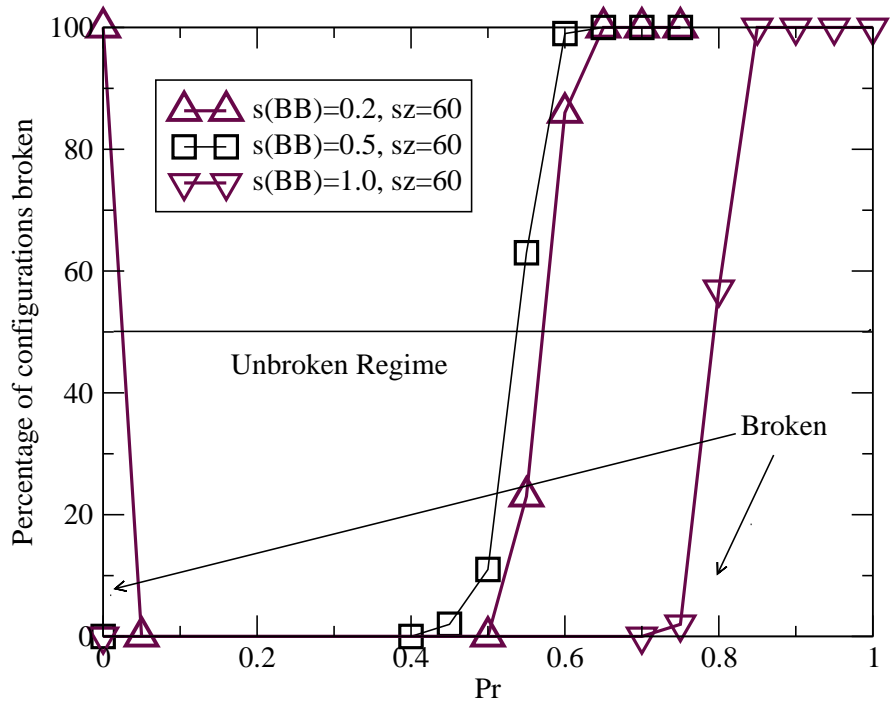}
\end{center}
\caption{The fraction of broken configurations as $s(BB)$ varies from 0.1 to 1.0 are shown as a function of $Pr$. For low values of $s(BB)$ there is a transition from broken $\rightarrow$  unbroken $\rightarrow$ broken, but for 0.5 and above there is only one transition from unbroken to broken regime as $Pr$ increases. } \label{brconf}
\end{figure}

Fig(\ref{brconf}) shows the number of configurations with spanning cracks as $s(AA)$ varies.

 \section{Discussion}
\begin{figure}[ht]
\begin{center}
\includegraphics[width=14.0cm, angle=0]{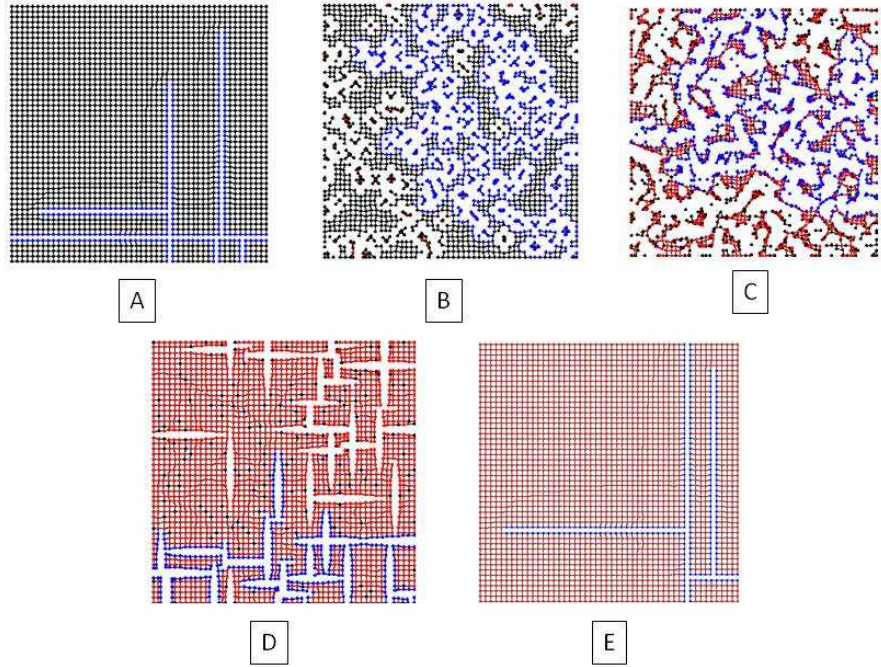}
\end{center}
\caption{'Snapshots' of simulations for $Pr$ = 0.0, 0.05, 0.5, 0.95 and 1.0. Blue color marks the spanning crack, which becomes more and more tortuous as $Pr$ deviates from 0 or 1. This explains why the energy to break the composite is more in figure(\ref{venergy}.} \label{energymax}
\end{figure}

 Comparing fig(\ref{venergy}) with fig(\ref{bondbr}), the low plateau region in fig(\ref{energymax}) corresponds to the region rich in B, where mainly B-B bonds break, but these are of low energy. Towards higher $Pr$ other bonds break in much larger numbers, thus a peak in energy forms. We have varied the thresholds only in this study, keeping spring constants fixed, other combinations can be tried out.
 
The challenge of producing better materials for practical applications by forming composites is
still being actively pursued \cite{zuda, silva}. The reason why composites are stronger, is due to cracks being arrested or deflected in composite media, whereas in homogeneous media they usually proceed along straight paths. 
 Recent simulation work reported on spring lattice models \cite{urabe} also show that meandering cracks in a composite imply a greater strength.
 Experimental and theoretical work on different aspects of crack formation through desiccation and mechanical stress is now a considerable volume of literature \cite{shor,kit,leung,nakahara,col2,born}. We have employed a model previously developed for desiccation cracks \cite{sadhu,jp1, epje}, but the incorporation of strain in the model is very general, so it is applicable to various situations.
 
 The first part of the study shows systems of composites (figure(\ref{energymax})), both of which are brittle in the pristine state, but require more energy to break when combined in a composite. The energy required for breaking is not equal to the strength, but is related to it, so we may consider this as a demonstration of increased strength of composites.
 
 The second part shows that a percolation threshold for spanning cracks, which break the system in two, depends on several factors. These include brittleness and ductility of the materials, represented here by the spring constants and thresholds. While percolating cracks have been studied and reviewed in \cite{nukala}, the effect in composites has not been reported. In the disordered media discussed previously spring breaking thresholds have finite distribution, but are are not bi-disperse as in the present work.
 
 We have not yet studied very large systems in order to identify the threshold values very precisely, at present we only show the existence of such thresholds. Studying different sizes and looking for finite size effects may be an interesting project for the future.
 
 In conclusion,we may say that the present study provides  some new angles to the problem of crack formation in complex materials, which may help in developing new materials.
 
 \section{Acknowledgment} The authors are very grateful to K.H. Hoffmann and Peter Blaudeck for helping with computer graphics.

\newpage

\newpage

\newpage

\newpage
\begin{figure}[ht]
\begin{center}
\includegraphics[width=14.0cm, angle=0]{bondbr.eps}
\end{center}
\caption{The number of different types of bonds broken and the total number of broken bonds is shown as function of $Pr$ for the parameters in figure\ref{venergy}}. The number of broken bonds is almost same whatever $s(BB)$, but the energy varies significantly. \label{bondbr}
\end{figure}

\newpage

\newpage

\newpage

\end{document}